\documentclass[twocolumn,aps,groupedaddress,floatfix,superscriptaddress]{revtex4-1}
\usepackage{ae}
\usepackage[utf8]{inputenc}
\usepackage{graphicx}
\usepackage{hyperref}
\usepackage[leqno]{amsmath}
\usepackage{amsfonts}
\usepackage{amssymb}
\usepackage{bm}
\usepackage{xcolor,colortbl}
\newcommand{\RM}[1]{\MakeUppercase{\romannumeral #1{}}} 

\newcommand{\sm}{\sigma^-}
\newcommand{\spl}{\sigma^+}

\begin{document}
\title{Integrated $^{9}$Be$^{+}$ multi-qubit gate device for the ion-trap quantum computer}

\author{H.~Hahn}
\affiliation{Physikalisch-Technische Bundesanstalt, Bundesallee 100, 38116 Braunschweig, Germany}
\affiliation{Institute of Quantum Optics, Leibniz University Hannover, Welfengarten 1, 30167 Hannover, Germany}
\author{G.~Zarantonello}
\affiliation{Physikalisch-Technische Bundesanstalt, Bundesallee 100, 38116 Braunschweig, Germany}
\affiliation{Institute of Quantum Optics, Leibniz University Hannover, Welfengarten 1, 30167 Hannover, Germany}
\author{M.~Schulte}
\affiliation{Institute for Theoretical Physics and Institute for Gravitational Physics (Albert-Einstein-Institute), Leibniz University Hannover, Appelstrasse 2, 30167 Hannover, Germany}
\author{A.~Bautista-Salvador}
\affiliation{Physikalisch-Technische Bundesanstalt, Bundesallee 100, 38116 Braunschweig, Germany}
\affiliation{Institute of Quantum Optics, Leibniz University Hannover, Welfengarten 1, 30167 Hannover, Germany}
\affiliation{Laboratory for Nano- and Quantum Engineering, Leibniz University Hannover, Schneiderberg 39, 30167 Hannover, Germany}
\author{K.~Hammerer}
\affiliation{Institute for Theoretical Physics and Institute for Gravitational Physics (Albert-Einstein-Institute), Leibniz University Hannover, Appelstrasse 2, 30167 Hannover, Germany}
\author{C.~Ospelkaus}
\email{christian.ospelkaus@iqo.uni-hannover.de}
\affiliation{Physikalisch-Technische Bundesanstalt, Bundesallee 100, 38116 Braunschweig, Germany}
\affiliation{Institute of Quantum Optics, Leibniz University Hannover, Welfengarten 1, 30167 Hannover, Germany}
\affiliation{Laboratory for Nano- and Quantum Engineering, Leibniz University Hannover, Schneiderberg 39, 30167 Hannover, Germany}

\begin{abstract}
We demonstrate the experimental realization of a two-qubit M{\o}lmer-S{\o}rensen gate on a magnetic field-insensitive hyperfine transition in $^9$Be$^+$ ions using microwave-near fields emitted by a single microwave conductor embedded in a surface-electrode ion trap. The design of the conductor was optimized to produce a high oscillating magnetic field gradient at the ion position. The measured gate fidelity is determined to be $98.2\pm1.2\,\%$ and is limited by technical imperfections, as is confirmed by a comprehensive numerical error analysis. The conductor design can potentially simplify the implementation of multi-qubit gates and represents a self-contained, scalable module for entangling gates within the quantum CCD architecture for an ion-trap quantum computer.
\end{abstract}

\maketitle
Following the proposal by Cirac \textit{et al.}~\cite{cirac_quantum_1995}, trapped atomic ions have shown to be a promising and pioneering platform for implementing elements of quantum information processing (QIP)~\cite{blatt_entangled_2008, monroe_scaling_2013}. Qubits are encoded in the internal states of individual ions and shared motional modes are used as a `quantum bus' for multi-qubit operations. Towards a large-scale universal quantum processor based on trapped-ion qubits, the `Quantum Charge-Coupled Device' (QCCD)~\cite{wineland_experimental_1998,kielpinski_architecture_2002} is considered a possible scalable hardware implementation. It relies on microfabricated multi-zone ion trap arrays in which quantum information is processed in dedicated zones interconnected via ion transport. While some key requirements such as high-fidelity ion transport~\cite{blakestad_high-fidelity_2009} and fault-tolerant single-qubit gates~\cite{brown_single-qubit-gate_2011, harty_high-fidelity_2014} have already been demonstrated in multiple setups, high-fidelity multi-qubit gates~\cite{gaebler_high-fidelity_2016,ballance_high-fidelity_2016} below the fault-tolerant threshold still remain challenging. In this context, entangling gates driven by microwave fields~\cite{mintert_ion-trap_2001,ospelkaus_trapped-ion_2008} represent a technically less demanding alternative to laser-induced gates as microwave signals can typically be controlled more easily than optical fields from highly specialized laser systems. The microwave approach avoids spontaneous scattering as a fundamental source of infidelities~\cite{ozeri_errors_2007} and experimental fidelities~\cite{ospelkaus_microwave_2011,khromova_designer_2012,weidt_trapped-ion_2016,harty_high-fidelity_2016} are approaching the fidelities of the best laser-driven gates~\cite{gaebler_high-fidelity_2016,ballance_high-fidelity_2016}. Here we focus on the near-field microwave~\cite{ospelkaus_trapped-ion_2008} gate approach, where the leading sources of infidelity in implementations so far comprise the spatio-temporal stability of the microwave near-field pattern~\cite{ospelkaus_microwave_2011,warring_techniques_2013} or fluctuating AC Zeeman shifts~\cite{harty_high-fidelity_2016,sepiol_high-fidelity_2016}. We note that in the latter work the error contribution arising from fluctuating AC Zeeman shifts has been reduced to $\lesssim0.1\,$\% through the use of a dynamical decoupling scheme~\cite{harty_high-fidelity_2016}.

In this letter, we realize a two-qubit gate using a tailored microwave conductor embedded in a surface-electrode trap optimized to produce high oscillating magnetic near-field gradients and low residual fields at the ion position, thus directly addressing the main sources of error in previous near-field gates. The gate is realized on a field-independent hyperfine qubit in $^9$Be$^+$ ions, a promising ion species for scalable QIP~\cite{home_complete_2009,tan_demonstration_2013,negnevitsky_repeated_2018-1}, following the M{\o}lmer-S{\o}rensen (MS)~\cite{sorensen_quantum_1999,molmer_multiparticle_1999,milburn_ion_2000,solano_deterministic_1999} protocol. The implementation is based on an optimized single-conductor design which can be thought of as the prototype of a scalable multi-qubit gate module for an ion-trap quantum computer based on surface-electrode trap arrays. The measured gate fidelity of $98.2\pm1.2\,\%$ is purely limited by technical imperfections, in agreement with a numerical analysis. 

The surface-electrode trap was fabricated at the PTB cleanroom facility employing the single-layer method as detailed in~\cite{bautista-salvador_multilayer_2019} on an AlN substrate~\cite{NoteX}. Gold electrodes are about $11\,\mathrm{\mu m}$ thick and separated by $5\,\mathrm{\mu m}$ gaps. Aiming to remove potential organic residuals on top of the electrode surfaces, the trap was cleaned in an \textit{ex-situ} dry-etching process before being installed in a UHV vacuum chamber at room-temperature. Electrical connectivity is provided by wire bonding to a printed-circuit board for DC signal filtering and signal routing. 

\begin{figure}[htbp]
	\centering
	\includegraphics[width=0.9\columnwidth]{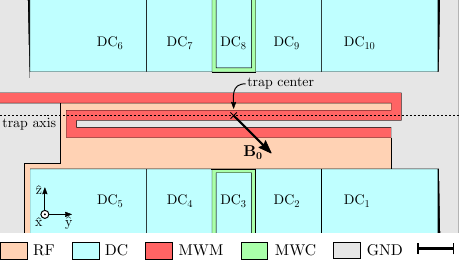}
	\caption{Electrode configuration around the trap center. The ions are confined $70\,\mu\mathrm{m}$ above the surface using 10 DC control electrodes and one split RF electrode. The microwave conductor labeled MWM produces an oscillating magnetic near-field gradient in the $\hat{x}$-$\hat{z}$ plane in order to drive an entangling gate on two hyperfine qubits. The two microwave conductors labeled MWC can each generate an oscillating magnetic field for global manipulation of the spin state in the $\mathrm{^{2}S_{1/2}}$ ma\-ni\-fold. Scale bar: $150\,\mu\mathrm{m}$ (bottom right)}
	\label{fig:figure1}
\end{figure}

Fig.~\ref{fig:figure1} shows a top view of the trap center. Besides the microwave conductor MWM, which produces a magnetic field quadrupole~\cite{wahnschaffe_single-ion_2017} suitable for driving multi-qubit gates, the electrode layout includes two microwave conductors labeled MWC to generate an oscillating magnetic field for global spin state manipulation. The ions are radially and axially confined at an ion-to-electrode distance of about $70\,\mathrm{\mu m}$ using one radio frequency (RF) and 10 DC control electrodes, respectively. With an applied RF voltage of $150\,\mathrm{V_{pp}}$ at $\Omega_{\mathrm{RF}}\simeq 2\pi \times 88.2\,\mathrm{MHz}$, the motional mode frequencies of the radial rocking modes of a two-ion $^{9}$Be$^{+}$ crystal were measured to be $(\omega_{\mathrm{r1}},\omega_{\mathrm{r2}}) \simeq 2\pi\times(6.275, 6.318)\,\mathrm{MHz}$.

\begin{figure}[htbp]
	\centering
	\vspace{0.5cm}
	\includegraphics[width=0.9\columnwidth]{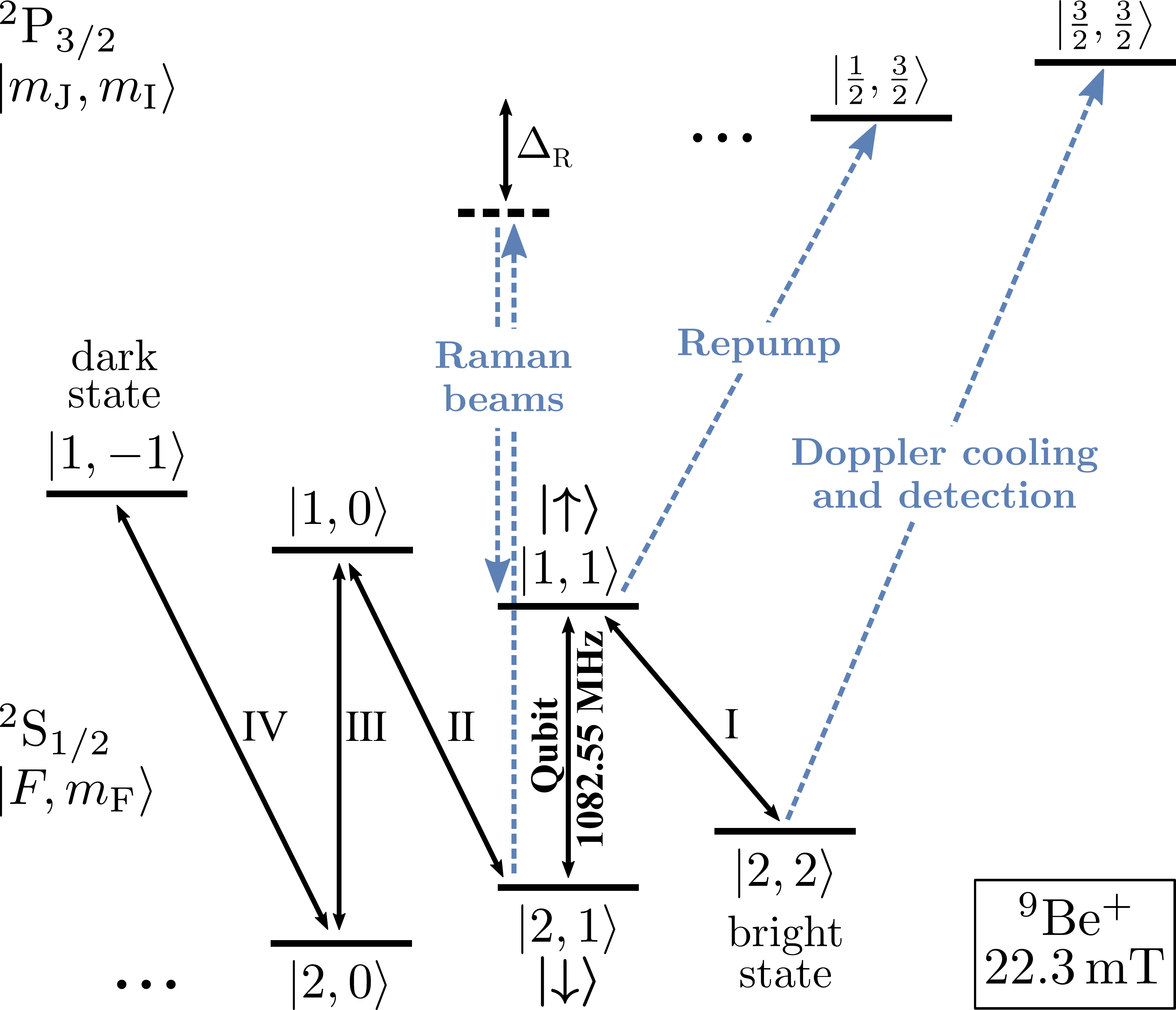}
	\caption{Relevant energy levels of $^9$Be$^{+}$ at $22.3\,\mathrm{mT}$. The transitions indicated as blue dashed lines are addressed by laser beams while the transitions indicated by black solid lines are driven by microwave radiation. The Raman beam detuning is $\Delta_{\mathrm{R}}\simeq100\,\mathrm{GHz}$ below the line center of the $\mathrm{^{2}P_{3/2}}$ manifold.}
	\label{fig:figure2}
\end{figure}

The static magnetic field $\bm{\mathrm{B_0}}$ defining the quantization axis at an angle of $45\,^{\circ}$ with respect to the trap axis is produced by a hybrid setup consisting of two permanent magnet assemblies and a pair of compensation coils~\cite{hakelberg_hybrid_2018}. At the ion position, this setup generates a magnetic field of $\left|\bm{\mathrm{B_0}}\right|=22.3\,\mathrm{mT}$ forming a first-order magnetic field insensitive qubit~\cite{langer_long-lived_2005} on the hyperfine levels $\mathrm{^{2}S_{1/2}}\left|F=1,\,m_{\mathrm{F}}=1\right>\equiv\left|\uparrow\right>$ and $\mathrm{^{2}S_{1/2}}\left|F=2,\,m_{\mathrm{F}}=1\right>\equiv\left|\downarrow\right>$ with an unperturbed transition frequency of $\omega_{\mathrm{0}}\simeq2\pi\times1082.55\,\mathrm{MHz}$, cf. Fig.~\ref{fig:figure2}. Here, $F$ and $m_{\mathrm{F}}$ represent the quantum numbers for the ion's total angular momentum and its projection on the quantization axis, respectively. 

Ions are loaded using laser ablation combined with photoionization~\cite{wahnschaffe_single-ion_2017}. State initialization is done via optical pumping to the $\left|2,\,2\right>$ state (also referred to as the bright state) and subsequent Doppler cooling on the closed-cycle transition $\mathrm{^{2}S_{1/2}}\left|2,2\right>$ $\leftrightarrow$ $\mathrm{^{2}P_{3/2}}\left|m_{\mathrm{J}}=\frac{3}{2},m_{\mathrm{I}}=\frac{3}{2}\right>$  (where $m_{\mathrm{J}}$ and $m_{\mathrm{I}}$ are the projections of the total electronic and nuclear angular momenta onto the quantization axis). Resolved sideband cooling is performed by a pair of counter-propagating Raman beams aligned along the $\hat{z}$ direction. Each sideband cooling cycle consists of a global $\pi$ rotation on the hyperfine transition labeled `\RM{1}' in Fig.~\ref{fig:figure2}, followed by an optical red sideband pulse on the qubit transition and a repumping sequence to transfer all population back to the initial bright state. The repumping sequence comprises multiple microwave induced $\pi$ rotations on the qubit transition and laser pulses on the $\mathrm{^{2}S_{1/2}}\left|1,1\right>$ $\leftrightarrow$ $\mathrm{^{2}P_{3/2}}\left|\frac{1}{2},\frac{3}{2}\right>$ transition similar to~\cite{warring_techniques_2013}. Employing interleaved sideband cooling sequences on both radial rocking modes of a two-ion crystal, we measure a mean phonon number $\bar{n}$ of the modes to be $\bar{n}_{\mathrm{r1}}\simeq0.27$ and $\bar{n}_{\mathrm{r2}}\simeq0.11$ assuming a thermal distribution. The heating rate was determined to be $\dot{\bar{n}}_{\mathrm{r2}}\simeq28\,\mathrm{s^{-1}}$. In an earlier measurement with similar radial mode frequencies, the heating rates of a single-ion's low-frequency (LF) and high-frequency (HF) radial mode were found to be about $\dot{\bar{n}}_{\mathrm{LF}}\simeq116\,\mathrm{s^{-1}}$ and $\dot{\bar{n}}_{\mathrm{HF}}\simeq122\,\mathrm{s^{-1}}$, respectively.

To perform a MS two-qubit entangling gate on the qubit transition using near-field microwaves~\cite{ospelkaus_trapped-ion_2008}, we subsequently initialize the qubits in $\left|\uparrow\uparrow\right>$ and apply a bichromatic microwave current to MWM at the frequencies $\omega_{\mathrm{RSB}}=\omega_{\mathrm{0}}+ \Delta - (\omega_{\mathrm{r2}}+\delta)$ and $\omega_{\mathrm{BSB}}=\omega_{\mathrm{0}}+\Delta + (\omega_{\mathrm{r2}}+\delta)$. Here $\delta$ is the gate detuning from the high frequency rocking mode of $N=2$ ions at $\omega_{\mathrm{r2}}$ and $\Delta$ is the differential AC Zeeman shift of the unperturbed qubit transition induced by the bichromatic field. In the ideal case, the implemented dynamics can be described by the Hamiltonian
\begin{equation}
	\label{eq:HMS}
	H_{\mathrm{MS}} = \frac{\Omega}{2}\sum_{j=1}^N (\sigma_j^+ + \sigma_j^-) (a_{r2} e^{\mathrm{i}\delta \, t} + a_{r2}^{\dagger} e^{-\mathrm{i}\delta \, t})\quad,
\end{equation}
where $\Omega$ is the gate Rabi frequency, $a_{r2}$ ($a_{r2}^{\dagger}$) represents the mode's annihilation (creation) operator and we define $\sigma^{\pm}=1/2(\sigma^{\mathrm{x}} \pm \mathrm{i} \sigma^{\mathrm{y}})$ with $\sigma^{\mathrm{x}}$ and $\sigma^{\mathrm{y}}$ being the Pauli matrices. Following~\cite{sorensen_entanglement_2000,sackett_experimental_2000}, we apply the interaction on $\left|\uparrow\uparrow\right>$ in order to produce the maximally entangled state $\left|\Psi\right>=1/\sqrt{2}(\left|\uparrow\uparrow\right>+\mathrm{i}\left|\downarrow\downarrow\right>)$ at time $\tau=\pi\sqrt{K}/\Omega$ (where $K$ is an integer number) and calculate the resulting state preparation fidelity $\mathcal{F}\equiv\left<\Psi\right|\rho\left|\Psi\right>=1/2(P_{\uparrow\uparrow} + P_{\downarrow\downarrow}) +\left| \rho_{\uparrow\uparrow,\downarrow\downarrow}\right|$ by determining the far off-diagonal element $\rho_{\uparrow\uparrow,\downarrow\downarrow}$ of the system's density matrix $\rho$ as well as the population probabilities in $\left|\downarrow\downarrow\right>$, $\left|\uparrow\downarrow\right>$ and $\left|\downarrow\uparrow\right>$, and $\left|\uparrow\uparrow\right>$ given by $P_{\downarrow\downarrow}$, $P_{\uparrow\downarrow,\downarrow\uparrow}$ and $P_{\uparrow\uparrow}$, respectively.

\begin{figure*}[t!]
	\centering
	\includegraphics[width=.95\textwidth]{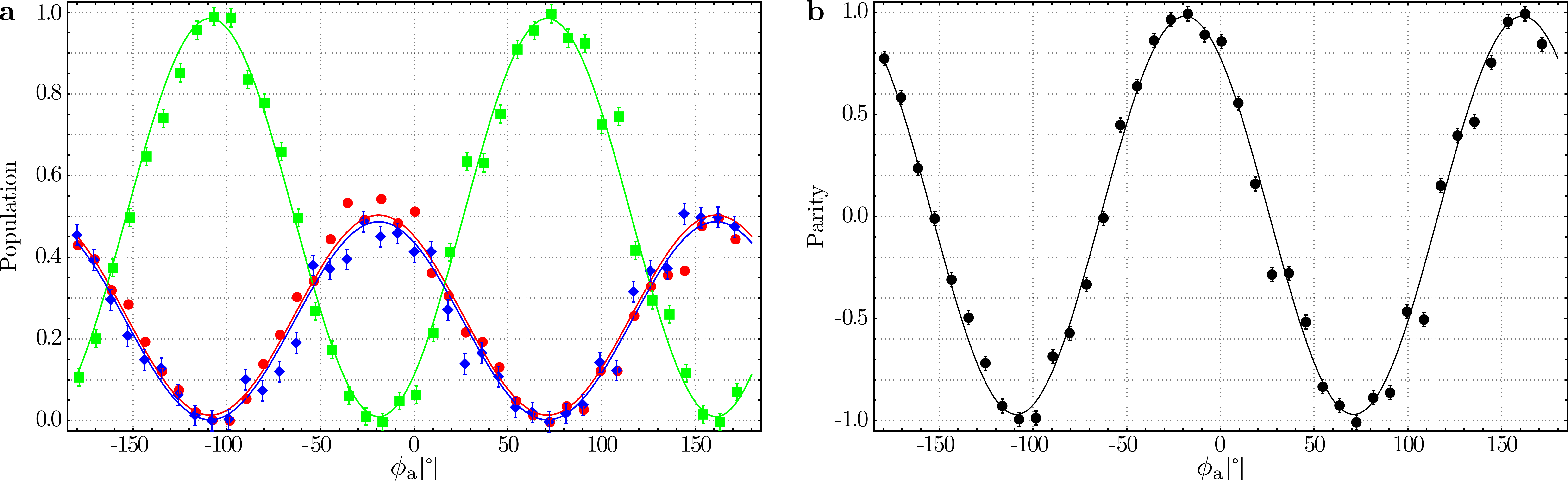}
	\caption{Population and parity oscillation as a function of the phase $\phi_{\mathrm{a}}$ of a $\pi/2$ analysis pulse applied after preparing the maximally entangled state 
	$\left|\Psi\right>=1/\sqrt{2}(\left|\uparrow\uparrow\right>+\mathrm{i}\left|\downarrow\downarrow\right>)$
	utilizing the near-field pattern of the MWM conductor. The solid lines are sinusoidal fits to the observed data while each data point represents the average of 200 experiments. 
	The error bars result from the fit of the weighted Poissonians to the experimental data.
	\textbf{a}, Population in $\left|\downarrow\downarrow\right>$ ($P_{\downarrow\downarrow}$; red circles),  $\left|\uparrow\downarrow\right>$ and $\left|\downarrow\uparrow\right>$ ($P_{\uparrow\downarrow,\downarrow\uparrow}$; green squares) and $\left|\uparrow\uparrow\right>$ ($P_{\uparrow\uparrow}$; blue diamonds). The fits yield  $P_{\uparrow\uparrow}+P_{\downarrow\downarrow}=0.990\pm0.021$. 
	\textbf{b}, Parity $\Pi(\phi_{\mathrm{a}})=P_{\downarrow\downarrow}(\phi_{\mathrm{a}})+P_{\uparrow\uparrow}(\phi_{\mathrm{a}})-P_{\uparrow\downarrow,\downarrow\uparrow}(\phi_{\mathrm{a}})$ oscillation with a fitted amplitude of $A_{\Pi}=0.975\pm0.012$. 
	}
	\label{fig:figure3}
\end{figure*}

Experimentally, we infer these quantities by adding a $\pi/2$ analysis pulse after the gate operation and observe the global ion fluorescence as a function of the analysis pulse's phase $\phi_{\mathrm{a}}$ employing fluorescence detection on the closed-cycle transition for $400\,\mathrm{\mu s}$. Prior to detection, we transfer the population in $\left|\uparrow\right>$ back to the bright state and shelve the population in $\left|\downarrow\right>$ to $\left|1,\,-1\right>$ (also referred to as the dark state) using a sequence of $\pi$ rotations on the transitions `\RM{1}' and `\RM{2}'-`\RM{4}', respectively. The populations are determined by repeating the experiment 200 times for each value of $\phi_{\mathrm{a}}$ and fitting the resulting histograms to a sum of weighted Poisson distributions representing 0, 1 and 2 ions bright. We calibrate the mean of the distributions to a reference two-ion Ramsey experiment which we assume to produce an ideal outcome following the same procedure of~\cite{ospelkaus_microwave_2011} (see online methods). To account for off-resonant optical pumping effects, we modify the three Poissonians to include all depumping processes from the shelved dark state to the bright state during the detection interval~\cite{langer_high_2006}. By consequence, we expect the resulting populations to compensate imperfect state preparation and detection within the present level of accuracy. Finally, $\rho_{\uparrow\uparrow,\downarrow\downarrow}$ can be deduced by calculating the parity $\Pi(\phi_{\mathrm{a}})=P_{\downarrow\downarrow}(\phi_{\mathrm{a}})+P_{\uparrow\uparrow}(\phi_{\mathrm{a}})-P_{\uparrow\downarrow+\downarrow\uparrow}(\phi_{\mathrm{a}})$ while $\phi_{\mathrm{a}}$ is varied and extracting the magnitude $\left|A_{\Pi}\right|$ of the parity oscillation equal to $\left| 2\,\rho_{\uparrow\uparrow,\downarrow\downarrow}\right|$~\cite{sackett_experimental_2000}. Fig.~\ref{fig:figure3} shows the population and parity signal after performing the two-qubit gate operation using the MWM conductor. From sinusoidal fits to the extracted populations (solid lines), we calculate a corresponding gate fidelity of $98.2\pm1.2\%$. The error on the fidelity is derived from the uncertainties in the population fits. In more detail, we apply a power of $\sim5.5\,\mathrm{W}$ to each sideband tone, resulting in gate Rabi frequency of $\Omega/2\pi=1.071\,\mathrm{kHz}$ and an inferred near-field gradient of around $19\,\mathrm{T\,m^{-1}}$. For a single sideband the corresponding residual magnetic field at the ion position is highly suppressed by the optimized conductor geometry, resulting in an on-resonance  Rabi frequency of about $\Omega_{\mathrm{c}}/2\pi\sim450\,\mathrm{kHz}$. In case the bichromatic gate drive is applied, the accompanied differential AC Zeeman shift is measured to be $\Delta/2\pi=4.37\,\mathrm{kHz}$ and is predominantly caused by $\sigma$-components of the residual magnetic fields as the induced shifts of the $\pi$-components mainly cancel each other assuming an equal power in both sideband tones. In order to suppress off-resonant spin excitations, we adiabatically shape the envelope of the gate pulse at its beginning and end with a $2\,\mathrm{\mu s}$ long cumulative error function using a microwave envelope stabilization circuit based on a fast analog multiplier and a digital PI controller~\cite{hannig_highly_2018} with the regulator setpoint generated by an arbitrary waveform generator~\cite{bowler_arbitrary_2013}. We optimize the gate by fixing the pulse duration to the expected value based on the gate Rabi frequency and subsequently scanning the gate detuning resulting in the highest fidelity. Following this procedure, we find an optimal gate time and detuning of $\tau=808\,\mathrm{\mu s}$ and $\delta/2\pi=3.4\,\mathrm{kHz}$, respectively, corresponding to $K=3$ loops in motional phase space. Here the mismatch to the theoretically predicted detuning at $\delta_{\mathrm{theory}}/2\pi=3.71\,\mathrm{kHz}$ was tracked back to a systematic frequency offset from an independent radial mode frequency measurement as well as a radial mode frequency `chirp' of $0.3\,\mathrm{Hz\,\mu s^{-1}}$ during the gate pulse. Qualitatively, a similar effect is also observed in other experiments using near-field gradients and appears to be inherent to warm-up processes in the microwave-generating structures~\cite{sepiol_high-fidelity_2016}. This hypothesis is supported by our observation of  a saturation behavior of the `chirp' at around $1\,\mathrm{ms}$. Consequently, we reduce the impact of the `chirp' by preceding the gate with a $400\,\mathrm{\mu s}$ long warm-up pulse with the duration chosen conservatively to avoid excessive heating of trap structures. 

In order to identify current infidelity contributions in producing the maximally entangled state, we simulate the dynamics of the system using a master equation considering experimentally determined input parameters. This becomes necessary as the exact propagator may no longer be obtained analytically in the presence of additional error sources. The master equation is given by 
\begin{equation}\label{eq:MEQ-Full}
	\dot{\rho} = -\mathrm{i} [H, \rho] + \mathcal{L}_h \rho + \mathcal{L}_d \rho
\end{equation}
where the Hamiltonian is $H = \tilde{H}_{\mathrm{MS}} + H_{\mathrm{m}} + H_{\mathrm{z}} + H_{\mathrm{spec}}$, and
\begin{equation}
  \tilde{H}_{\mathrm{MS}} = \dfrac{1}{2} \sum_{j=1}^N (\Omega^{\mathrm{B}} (t)\spl_j a_{r2}^{\dagger} e^{-\mathrm{i} \delta t} + \Omega^{\mathrm{R}} (t) \spl_j a_{r2} e^{\mathrm{i}\delta t} ) + \mathrm{H.c.}
\end{equation}
is an extension to the ideal case presented in Eq.(\ref{eq:HMS}). Here we have assumed equal Rabi frequencies and phases for both ions, which is true in the experiment to the best of our knowledge. Further, $ H_{\mathrm{m}} = \delta_{\epsilon} (t) a_{r2}^{\dagger}a_{r2}$ describes the instability of the rocking mode frequency, $ H_{\mathrm{z}}= \Delta_{\epsilon} (t)/2 \sum_j \sigma^{\mathrm{z}}_j $ gives the uncompensated AC Zeeman shift resulting from shot-to-shot microwave power fluctuations, general $\Omega^{\mathrm{B}}$ and $\Omega^{\mathrm{R}}$ allow an imbalance in the two sideband Rabi frequencies, and time dependencies of the pulse shape are taken into account by the time dependent parameters. Couplings via additional, off-resonant motional modes are included by the term $ H_{\mathrm{spec}} = \Omega_{r1}/2 \sum_{j=1}^N (\spl_j + \sm_j) (a_{r1} e^{\mathrm{i}(\Delta \nu +\delta) t} + a_{r1}^{\dagger} e^{-\mathrm{i}(\Delta \nu +\delta) t})$ whereby we limit ourselves to the nearest mode only (with $\Omega_{r1} \simeq \Omega$), which contributes the largest error of this kind. In addition to the unitary dynamics, motional heating to a thermal state with $\overline{n}_{\mathrm{th}} \gg 1$ and qubit decoherence are considered by the Lindblad terms~\cite{sorensen_entanglement_2000} $\mathcal{L}_{\mathrm{h}} \rho = \gamma_{\mathrm{h}} (\mathcal{D}[a_{r2}]\rho + \mathcal{D}[a_{r2}^{\dagger}]\rho)$ with the heating rate $\gamma_{\mathrm{h}}$ in phonons per second and $\mathcal{L}_{\mathrm{d}} \rho = \gamma_{\mathrm{d}}/2 \sum_{j} \mathcal{D}[\sigma^{\mathrm{z}}_j]\rho$ with the decoherence rate $\gamma_{\mathrm{d}}$ respectively, where $\mathcal{D}[\hat{O}]\rho = \hat{O} \rho \hat{O}^{\dagger} - \hat{O}^{\dagger} \hat{O} \rho/2 - \rho \hat{O}^{\dagger} \hat{O}/2$.

Table~\ref{tab:ErrorBudget} lists contributions of the different error sources to the infidelity $1 - \mathcal{F}$. These values result from numerical simulations of the quantum dynamics according to Eq.~\eqref{eq:MEQ-Full} considering the ideal gate dynamics with addition of the corresponding noise in the form we described above. All simulations were done with QuTiP~\cite{noauthor_qutip:_2016} and used a truncated Hilbert space for the motional mode. For our analysis, including the first 25 Fock states was sufficient to reach convergence given the low initial thermal distribution and the small motional displacements during the gate.

\begin{table}[t]
  \centering
  \setlength\extrarowheight{4pt}
  \begin{tabular}{ p{2.7cm}  c  c }
    \hline
	Effect            & Parameter                                                                      & Infidelity \\
	[4pt]
	\hline
	Mode instability  & $\sqrt{\langle (\delta_{\epsilon}/\delta)^2\rangle} = 1.1\times 10^{-2}$       & $1.3\times10^{-2}$ \\
                      & $0.3\,\mathrm{Hz}\,\mu\mathrm{s}^{-1}$ `chirp' for $600\,\mu\mathrm{s}$        & \\
	[4pt]
	\rowcolor{gray!20}[2.5pt]
	Spectator mode    & $\Delta \nu = 2 \pi \times 42.5\,\mathrm{kHz}$                                 & $5.2\times 10^{-3}$ \\
	\rowcolor{gray!20}[2.5pt]  
	                  & with $\bar{n}_{\mathrm{r1}}=0.27$                                              & \\
	[4pt]
	Motional heating  & $\gamma_{\mathrm{h}} = \dot{\bar{n}}_{\mathrm{r2}} = 28\,\mathrm{s^{-1}} $     & $3.8\times 10^{-3}$ \\
	[4pt]
	\rowcolor{gray!20}[2.5pt]
	Off-resonant      & - (measured infidelity                                                         & $< 2.3 \times 10^{-3}$ \\
	[-0pt]
	\rowcolor{gray!20}[2.5pt]
	scattering loss	  & following~\cite{ballance_high-fidelity_2017,sepiol_high-fidelity_2016})        & \\
	[4pt]
	Qubit decoherence & $\tau_{\mathrm{d}} = 1/\gamma_{\mathrm{d}} > 0.5\,\mathrm{s} $                 & $< 9.3\times 10^{-4}$ \\
	[4pt]
	\rowcolor{gray!20}[2.5pt]
	Pulse shape       & see main text                                                                  & $6.3 \times 10^{-4}$ \\
	[4pt]
	ACZS fluctuations & $\sqrt{\langle (\Delta_{\epsilon}/\Delta)^2\rangle} = 8\times 10^{-4}$         & $1.1\times10^{-4}$ \\
	[4pt]
	\rowcolor{gray!20}[2.5pt]
	Rabi frequency 
	\newline 
	imbalance         & $\dfrac{\Omega_{\mathrm{R}} - \Omega_{\mathrm{B}}}{\Omega_{\mathrm{B}}} = 2.33\times10^{-2}$ & $4.1\times 10^{-6}$ 
	\\
	[4pt] 
  \end{tabular}
  \caption{Infidelity contributions from different sources of imperfections. The infidelity values result from numerical simulations of the quantum dynamics according to Eq.\eqref{eq:MEQ-Full} including the respective noise effect with a strength given by the measured parameter specified in the second column.}
  \label{tab:ErrorBudget}
\end{table}

\begin{figure}[htbp]
	\centering
	\vspace{0.5cm}
	\includegraphics[width=\columnwidth]{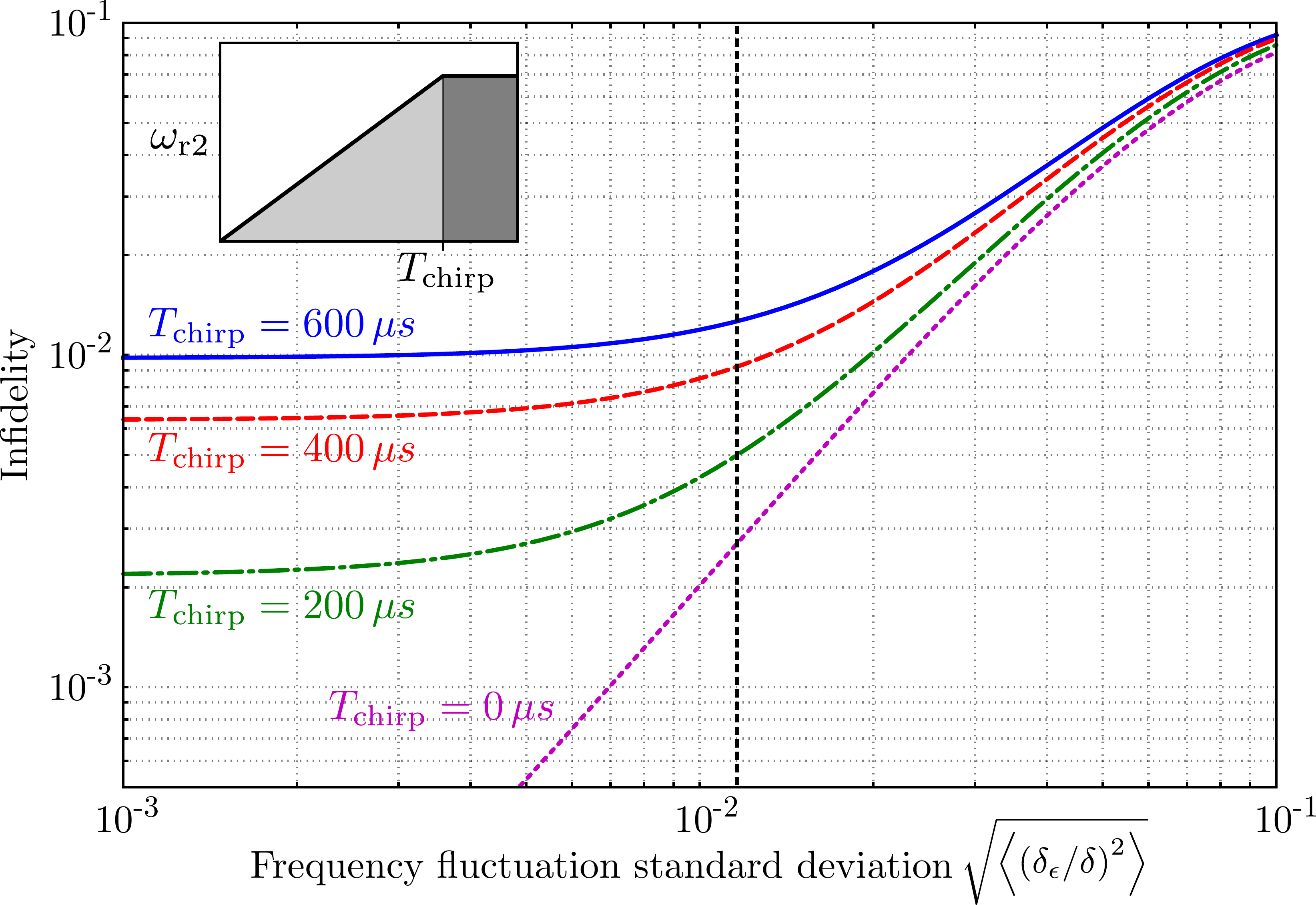}
	\caption{Simulated infidelity assuming different variations of the relative motional mode frequency and lengths of the frequency `chirp' while considering otherwise ideal gate dynamics. In all cases the gate parameters are as specified in the text. The experimental condition of the presented two-qubit gate is given by the intersection of the vertical line with the solid blue line. The inset illustrates the observed linear frequency increase of the selected motional mode of about $0.3\,\mathrm{Hz\,\mu s^{-1}}$ at the beginning of the gate pulse until time $T_{\mathrm{chirp}}$.}
	\label{fig:figure4}
\end{figure}

We examined the following effects, which we considered to be the most relevant, in more detail: the largest error according to our investigation results from the frequency instability of the rocking-mode which establishes the gate dynamics. This effect consists of two parts. On the one hand, normally distributed variations of the frequency with a standard deviation of $\sqrt{\langle (\delta_{\epsilon}/\delta)^2\rangle} = 1.1\times 10^{-2}$, inferred from a measured instantaneous linewidth of $ 2\pi \times 101\,\mathrm{Hz} $ in a calibration scan directly before the gate measurement. On the other hand, a frequency `chirp' within each gate that we model by a linear increase of $0.3\, \mathrm{Hz}\,\mu\mathrm{s}^{-1}$ within the first $600\,\mu\mathrm{s}$ and subsequent constant frequency leading to in total $1.3\,\%$ infidelity. While the mode fluctuations can be reduced by actively stabilizing the amplitude and frequency of the trap RF signal~\cite{johnson_active_2016,harty_high-fidelity_2013}, the `chirp' can be reduced by e.g. longer warm-up pulses. Simulations (see Fig.~\ref{fig:figure4}) also us to identify the individual contribution of each effect in view of further improvements.

The second largest contribution is caused by driving the identical spin-spin interaction via the additional low frequency rocking mode. The measured frequency spacing was $ \Delta \nu = \omega_{\mathrm{r2}}-\omega_{\mathrm{r1}} = 2 \pi \times 42.5\,\mathrm{kHz} $ and ground state cooling to $\bar{n}_{\mathrm{r1}} = 0.27$ was applied, resulting in an infidelity of $5.2\times10^{-3}$ from the competing gate dynamics. As this effect scales inversely with the radial mode splitting, it can be suppressed by engineering the trap potentials or suitable pulse sequences~\cite{choi_optimal_2014,milne_phase-modulated_2018}. Heating of the motion and decoherence of the qubits contribute $\sim0.1\%$ and $< 0.1\%$ infidelity, respectively. Again, both effects do not represent a fundamental limit to the gate performance and can be improved experimentally~\cite{hite_100-fold_2012, langer_long-lived_2005}. Off-resonant scattering on carrier transitions can lead to undesired excitations inside and outside the qubit-manifold and thus contribute a gate error. Here, an excitation other than on the qubit transition is much less probable due to the higher frequency difference of the driving field which is $>150\,\mathrm{MHz}$ detuned from the next spectator transition. Direct simulation of this effect was not performed due to the vastly different time-scales of the gate dynamics ($\sim\mathrm{kHz}$) and the carrier processes ($\sim \mathrm{GHz}$) which would have considerably increased the runtime of the numerical simulations. We instead performed direct measurements on a single qubit~\cite{ballance_high-fidelity_2017, sepiol_high-fidelity_2016} to evaluate the extent of this error which is then quantified to be $< 2.3\times10^{-3}$. Infidelities below $6.3\times10^{-4}$ resulted from distortion of the pulse shape, whereby we combine here the influence of adiabatic switching on and off as well as small changes of the Rabi frequency and AC Zeeman shift during the pulses which result from power transients on the ideally rectangular signal. Stabilization of the microwave power allowed to reduce the shot-to-shot fluctuations of the power, and accordingly of the AC Zeeman shift, to an extent that the simulated infidelity of $1.1\times10^{-4}$ contributes only insignificantly. The same applies to the imbalance of Rabi frequencies, cf. Table~\ref{tab:ErrorBudget}.

In conclusion, we have demonstrated a microwave-driven two-qubit gate with $^{9}$Be$^{+}$ ions using a single microwave conductor with an optimized design embedded in a surface-electrode ion trap. The design of the MWM conductor has been developed to generate a high magnetic near-field gradient with low residual field at the ion position, thus suppressing AC Zeeman shift fluctuations, an inherent error source of the near-field approach, to a simulated infidelity contribution of $\sim10^{-4}$. In contrast, according to the presented error budget, the main contributions can all be decreased upon technical improvements; the by far biggest of these ($1.3\,\%$) is consistent with the measured two-qubit gate infidelity of $1.8\pm1.2\,\%$. In addition to technical modifications to the apparatus, more elaborate gate schemes employing Walsh modulation~\cite{hayes_coherent_2012} or continuous dynamic decoupling~\cite{harty_high-fidelity_2016} can be applied in order to increase the gate fidelity as required for fault-tolerant quantum computation. In the future, the MWM conductor design can be used as an entangling gate unit of a `QCCD' architecture purely employing microwave-driven quantum gates. Moreover, the conductor design can also be integrated into a scalable multilayer trap~\cite{hahn_multilayer_2018,bautista-salvador_multilayer_2019}.

\subsection*{Data availability}
\noindent The datasets generated during and/or analysed during the current study are available from C. Ospelkaus (christian.ospelkaus@iqo.uni-hannover.de) on reasonable request.

\subsection*{Acknowledgements}
We acknowledge funding from DFG through CRC 1227 DQ-\textit{mat}, projects A01 and A06, and the clusters of excellence `QUEST' and `Quantum Frontiers', from the EU QT flagship project `MicroQC' and from PTB and LUH.   

\subsection*{Competing interests}
\noindent The authors declare that there are no competing interests.

\subsection*{Author contributions}
\noindent HH and GZ performed the measurements and analyzed the data. ABS produced the ion trap. MS and KH contributed the numerical simulations. CO devised the experiment plan. All authors participated in the error analysis and the realization of the manuscript.


\begin{thebibliography}{10}
\bibitem{NoteX}
	\bibinfo{note}{For the present trap, we chose the single-layer process in order
		to quickly test improvements that were made to the setup and trap orientation
		compared to ref. [45].}
\end{thebibliography}
\end{document}